\documentclass[fleqn,usenatbib]{mnras}

\usepackage{newtxtext,newtxmath}
\usepackage[T1]{fontenc}

\DeclareRobustCommand{\VAN}[3]{#2}
\let\VANthebibliography\thebibliography
\def\thebibliography{\DeclareRobustCommand{\VAN}[3]{##3}\VANthebibliography}

\usepackage{graphicx}	% Including figure files
\usepackage{amsmath}	% Advanced maths commands
\title[Spin distribution of BBHs in open cluster]{Spin distribution of binary black holes formed in open clusters}

\author[J. Kumamoto et al.]{
Jun Kumamoto,$^{1}$ \thanks{E-mail:kumamoto@astron.s.u-tokyo.ac.jp}
Michiko S. Fujii,$^{1}$ 
Alessandro A. Trani,$^{2}$ 
and Ataru Tanikawa$^{2}$
\\
$^{1}$ Department of Astronomy, Graduate School of Science, The University of Tokyo, 7-3-1 Hongo, Bunkyo-ku, Tokyo 113-0033, Japan\\
$^{2}$ Department of Earth Science and Astronomy, College of Arts and Sciences, The University of Tokyo, 3-8-1 Komaba, Meguro-ku, Tokyo 153-8902, Japan}

\date{Accepted XXX. Received YYY; in original form ZZZ}

\pubyear{2020}

\begin{document}
\label{firstpage}
\pagerange{\pageref{firstpage}--\pageref{lastpage}}
\maketitle

\begin{abstract}
We performed direct N-body simulations of open clusters with four
different metallicities. To investigate the effective
  spins of merging binary black holes (BBHs) originated from
these open clusters, we calculated the spin evolution of
  Wolf-Rayet (WR) stars with close companion stars (BBH progenitors),
  taking into account stellar wind mass loss and tidal spin-up of the
  WR stars. We found that BBH progenitors with smaller semi-major axes
  evolve to merging BBHs with greater effective spins because of
  strong tidal forces. In the local Universe, about 16\% of merging BBHs get
  effective spins larger than 0.1 even if BHs and their progenitors do
  not get spin angular momenta other than tidal forces exerted by
  their companion stars. If we assume that WR stars have flat and isotropic
  distribution of dimensionless spins just after common envelope
  phases, the effective spin distribution of merging BBHs is
similar to that inferred from gravitational wave
observations with LIGO and Virgo.
\end{abstract}


%%%%%%%%%%%%%%%%%%%%%%%%%%%%%%%%%%%%%%%%%%%%%%%%%%

%%%%%%%%%%%%%%%%% BODY OF PAPER %%%%%%%%%%%%%%%%%%

\section{Introduction}
\label{sec:intro}

The recent detections of gravitational waves with LIGO and Virgo have
revealed many mergers of two stellar-mass black holes (BHs)
\citep[e.g.][]{2019PhRvX...9c1040A, 2020arXiv201014527A}. Prior to their detection, it was not expected that there would be
so many massive (${>} 15 \rm\,M_\odot$) merging binary BHs (BBHs). Numerous scenarios have been suggested to
explain the formation process of these BBHs, but their origin is still unclear.

One of the main scenarios is via common envelope and mass transfer
evolution of isolated field binaries
\citep[e.g.][]{1973NInfo..27....3T, 1998ApJ...506..780B,
  2012ApJ...759...52D, 2014MNRAS.442.2963K, 2016Natur.534..512B,
  2020A&A...636A.104B, 2018MNRAS.474.2959G, 2020A&A...635A..97B,
  2020arXiv200801890T, 2020arXiv201007616T}. The other major scenarios
are the dynamical formation due to three-body encounters in
globular clusters \citep[e.g.][]{2000ApJ...528L..17P,
  2006ApJ...637..937O, 2008ApJ...676.1162S, 2010MNRAS.407.1946D,
  2011MNRAS.416..133D, 2010MNRAS.402..371B, 2013MNRAS.435.1358T,
  2014MNRAS.440.2714B, 2015PhRvL.115e1101R, 2016PhRvD..93h4029R,
  2017PASJ...69...94F, 2017MNRAS.469.4665P, 2017MNRAS.464L..36A,
  2018MNRAS.480.5645H, 2019ApJ...871...91Z, 2020MNRAS.492.2936A,
  2020ApJS..247...48K, 2021arXiv210109283W}, open cluster (or young
star cluster) \citep[e.g.][]{2014MNRAS.441.3703Z, 2014ApJ...781...81G,
  2016MNRAS.459.3432M, 2017MNRAS.467..524B, 2018MNRAS.473..909B,
  2018MNRAS.481.5123B, 2019MNRAS.483.1233R, 2019MNRAS.487.2947D,
  2019MNRAS.486.3942K, 2020MNRAS.495.4268K, 2019ApJ...886...25B,
  2021MNRAS.500.3002B, 2021arXiv210201689T} or galactic nuclei
\citep[e.g.][]{2009MNRAS.395.2127O, 2016ApJ...831..187A,
  2018ApJ...856..140H, 2018ApJ...865....2H, 2018ApJ...866...66M, 2019ApJ...875L..31H, 2019MNRAS.483..152A,
  2020ApJ...898...25T, 2020IAUS..351..174T, 2020ApJ...891...47A}. However, the
  origin of merging BBHs has been under debate.

In \citet[][hereafter Paper I]{2019MNRAS.486.3942K}, we performed
N-body simulations for open clusters with half-mass
densities of $10^4M_{\odot}~{\rm pc^{-3}}$, and found
a new channel for formation of BBHs merging within
the Hubble time. In open clusters, despite the short lifetimes of
massive main-sequence (MS) stars, they can form binaries due
to three-body interaction before they evolve to BHs because
the core-collapse time of open cluster is shorter than the lifetime of
massive MS stars. These binaries evolve to BBHs
merging within the Hubble time via common envelope evolution. In
\citet[][hereafter Paper II]{2020MNRAS.495.4268K}, we estimated the
local merger rate density of BBHs formed in open clusters to be $\sim
70~{\rm yr}^{-1} {\rm Gpc}^{-3}$ using the results of our $N$-body
simulations of open clusters with four different metallicity
models. This result is comparable to the merger rate density expected
from the gravitational wave observation with LIGO and Virgo
\citep{2019ApJ...882L..24A}.

The effective spin, $\chi_{\rm eff}$, is one of important
parameters for discussing the formation scenario of BBHs. The
distribution of $\chi_{\rm eff}$ estimated from the detection of
gravitational waves show that most of BBHs have small positive and
negative $\chi_{\rm eff}$ \citep{2019PhRvX...9c1040A,
  2020arXiv201014533T}.

One possible scenario for the origin of BH spin is tidal
spin-up of Wolf-Rayet (WR) stars before they evolve into
BHs. Several previous studies showed that WR stars
can be spun up if they have very close companion stars
\citep{2017MNRAS.467.2146K, 2017ApJ...842..111H, 2020ApJ...892...64P,
  2020A&A...636A.104B}.

In this paper, we investigate the effective spins of BBHs
formed in open clusters. For the members of BBHs formed in our
simulations, we calculate the spin evolution
of WR stars before they evolve into BHs. We find that even if
we assume that WR stars have zero spin angular momenta just after
common envelope evolution,
some BBH progenitors with small semi-major axes
evolve to BBHs with $\chi_{\rm eff}$ larger than
0.1. In subsolar-metallicity cases, some BBHs formed in
more metal-rich clusters have lower masses and higher
$\chi_{\rm eff}$. In the solar-metallicity case, our
simulations show that there is no BBH with $\chi_{\rm eff}$ larger
than 0.1.

The structure of this paper is as follows. Our simulation methods and
models are described in Section 2.  We show our spin evolution
model of WR stars (BH progenitors) to calculate effective spins of
BBHs in Section 3.
In Section 4, we investigate and discuss the distribution of the
effective spins of BBHs formed in our open cluster
simulations. We make a conclusion in Section 5.

%%%%%%%%%%%%%%%%%%%%%%%%%%%%%%%%%%%%%%%%%%%%%%%%%%
%%%%%%%%%%%%%%%%%%%%%%%%%%%%%%%%%%%%%%%%%%%%%%%%%%

\section{Simulations}
\label{sec:simulations}

We analyze the results of $N$-body simulations for open
clusters with different four metallicities
($Z=0.002, 0.005, 0.01, 0.02$) performed in Paper II. A summary of
these simulations is following.

\subsection{Initial conditions}
We have set up four cluster models. Table \ref{tab:models} gives an
overview of the models. The initial cluster mass ($M_{\rm cl,ini}$) is
$2500~M_{\odot}$. The number of runs ($N_{\rm run}$) depends
on metallicity. Since we expect that more metal-rich clusters
  have a smaller number of heavy BHs due to the strong stellar wind
  mass loss, we performed a larger number of runs for more metal-rich
  clusters for statistics (see also Section
  \ref{sec:StellarEvolution}).
%% For metal-rich models, we expect less heavy BHs due to the strong
%% stellar wind mass loss, we performed more runs (see also subsection
%% \ref{sec:StellarEvolution}).

\begin{table}
 \centering
 \caption{Models.}
 \label{tab:models}
 \begin{tabular}{lccc}
  \hline
               & $M_{\rm cl,ini} [M_\odot]$ & $Z$   & $N_{\rm run}$ \\
  \hline
  Model Z0002  & $2.5\times10^3$            & 0.002 & 360  \\
  Model Z0005  & $2.5\times10^3$            & 0.005 & 500  \\
  Model Z001   & $2.5\times10^3$            & 0.01  & 1000 \\
  Model Z002   & $2.5\times10^3$            & 0.02  & 1000 \\
  \hline
 \end{tabular}
\end{table}

We adopt the Plummer profile as the initial
density distribution of our open cluster
models \citep{1911MNRAS..71..460P};
\begin{equation}
    \rho(r) = \frac{3M_{\rm cl,ini}}{4\pi r_p^3} \left( 1+\frac{r^2}{r_p^2} \right)^{-5/2},
	\label{eq:plummer}
\end{equation}
\begin{equation}
    r_p = (2^{2/3}-1)^{1/2} r_{\rm hm},
	\label{eq:r_p}
\end{equation}
where $r_{\rm hm}$ is a half-mass radius. In all our models,
the half-mass radii are set to be 0.31\,pc. In this
case, the initial half-mass densities ($\rho_{\rm hm} =
3M_{\rm cl,ini}/8\pi r_{\rm hm}^3$) are equal to $10^4
M_{\odot}~{\rm pc}^{-3}$. This value is higher than half-mass
densities of currently observed open clusters. However, the initial density of star clusters was likely higher than
the currently observed one
\citep{2010MNRAS.406.2000M,2010ARA&A..48..431P}. The observed mass and half-mass
density of young massive clusters are typically $\sim10^4 M_{\odot}$ and
$10^{1-4} M_{\odot}~{\rm pc}^{-3}$, respectively \citep[see
  Figure 2 of][]{2010ARA&A..48..431P}. Among these clusters, younger
clusters tend to have higher densities, and the
ages of youngest clusters are about $2~{\rm Myr}$,
which is younger than the lifetime of massive MS stars. The
half-mass densities of these youngest clusters are
about $10^4 M_{\odot}~{\rm pc}^{-3}$. Therefore, we use this value for
initial half-mass density.

As the initial stellar mass function, we adopted the \citet{Kroupa2001} initial mass
function with a lower mass limit
of $m_{\rm min} = 0.08 M_{\odot}$ and an upper mass
limit of $m_{\rm max} = 150M_{\odot}$. This leads to an
average stellar mass of $\langle m \rangle =
0.586M_{\odot}$. Thus, the initial number of particles $N_{\rm ini} =
M_{\rm cl,ini}/\langle m \rangle$ is 4266.

In our model, the half-mass relaxation time is calculated with
the following equation:
\begin{equation}
    t_{\rm rh} \sim 0.711\frac{N}{\log(0.4N)}\left(\frac{\rho_{\rm hm}}{M_{\odot}{\rm pc^{-3}}}\right)^{-0.5}~{\rm Myr}.
	\label{eq:t_rh}    
\end{equation}
The core-collapse time is proportional to the relaxation
time, and the proportional factor depends on the maximum and average
masses of stars in the system. In our models, we obtain
\begin{equation}
  t_{\rm cc} \sim 0.07 t_{\rm rh,ini} \sim 0.7~{\rm Myr}
\end{equation}
\citep{2004ApJ...604..632G,2014MNRAS.439.1003F}. Since most
of BBHs are formed within a few hundred Myr, tidal disruptions of the
cluster due to the external tidal field are not significant in this short
period. Therefore, we do not assume any external tidal force in our simulations.

We do not assume primordial binaries. In Paper I,
we argued that primordial binaries do not significantly affect
the merger rate of BBHs formed in open clusters. This is because merging BBHs (or their progenitors) are dynamically formed even if there is no primordial binary. In addition, previous studies have shown that the number of BBH mergers from stellar clusters per solar mass unit is about $10^{-4}$--$10^{-6} \rm\,M_\odot^{-1}$, independently of metallicity or primordial binary fraction \citep[e.g.][]{2014MNRAS.441.3703Z,2019MNRAS.487.2947D,2020MNRAS.495.4268K}. Given the modest mass (${\sim}$2500 M$_\odot$) of open clusters, primordial binaries can not substantially increase the production rate of BBHs in open clusters.

\subsection{N-body simulations}

We use a direct $N$-body simulation code, {\tt NBODY6++GPU} \citep{Wang+2015}, which is an MPI-parallelised and GPU enabled version of {\tt NBODY6} \citep{1999PASP..111.1333A}. The motions of individual stars are calculated using a fourth-order Hermite scheme \citep{1992PASJ...44..141M}. The closer binaries have time steps much smaller than the time-scale of the evolution of open clusters. Such closer binaries are integrated using KS regularization \citep{KustaanheimoStiefel1965, 1993CeMDA..57..439M}.

Our simulations are performed using GPU cluster SGI Rackable C1102-GP8 (Reedbush-L) in the Information Technology Center, The University of Tokyo.

\subsection{Stellar evolution}
\label{sec:StellarEvolution}

The evolution of stellar radius, mass, and luminosity are
calculated using a stellar evolution model, {\tt SSE}
\citep{Hurley+2000}. We use updated mass loss model
\citep{Belczynski+2010}, which is contained in the latest version of
      {\tt NBODY6}. For binary evolution, {\tt NBODY6++GPU}
      contains a binary evolution model \citep{1997MNRAS.291..732T},
      which is an algorithm for rapid evolution binary star following
      the common envelope and mass transfer. We set a common envelope
      efficiency parameter, $\alpha$, to be $1/3$. The BH
      mass also strongly depends on metallicity. More metal-rich stars
      evolve into less massive BHs because of the stronger
      stellar wind (See Figure 1. of Paper II).

The natal kicks caused by asymmetric supernovae explosion are not
included in our simulation for simplification. The natal kicks may
affect to the formation rate of BBHs \citep{2013MNRAS.435.1358T},
however, most BHs are retained in clusters
\citep{2013ApJ...763L..15M}. Even if a fraction of BHs are ejected
  from star clusters due to the natal kicks, BBHs are hardened not
  only by the rest of BHs but also the other massive stars.

%%%%%%%%%%%%%%%%%%%%%%%%%%%%%%%%%%%%%%%%%%%%%%%%%%
%%%%%%%%%%%%%%%%%%%%%%%%%%%%%%%%%%%%%%%%%%%%%%%%%%

\section{Spin evolution model}
\label{sec:Spins}

\begin{figure}
	\includegraphics[width=\columnwidth]{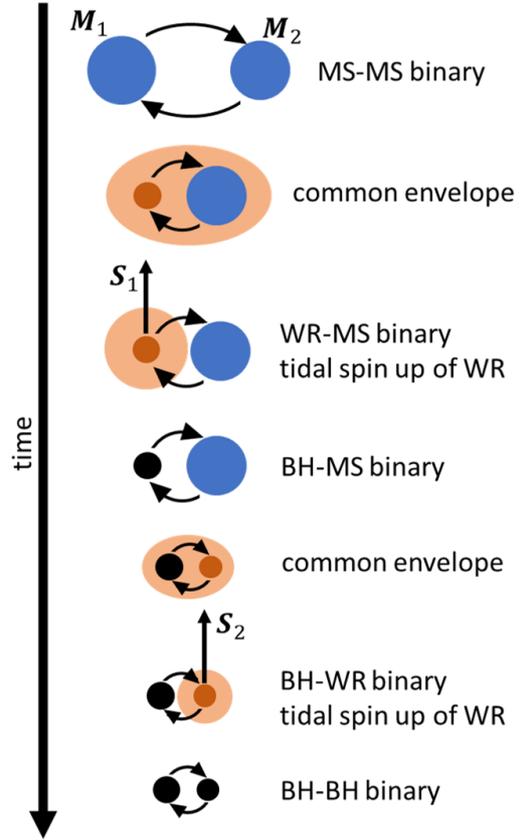}
    \caption{Schematic diagram of BBH formation with spinning BHs.}
    \label{fig:SpinForm}
\end{figure}

We describe here how BHs get their spin angular momenta in
  our model. Figure \ref{fig:SpinForm} shows the schematic diagram of
  BBH formation with spinning BHs. Two massive stars form a binary
  star through dynamical interactions. The primary star evolves to a
  post MS star, and fills its Roche lobe. Then, the primary star
  drives common envelope evolution, loses its envelope, and becomes a
  WR star. The WR star is tidally spun up by the secondary star, and
  collapses to a spinning BH. The secondary star can evolve to a
  spinning BH through the same mechanism. BH spins are higher, and
  more aligned to binary angular momentum vector as binary separations
  are smaller just after common envelope evolution. Note that BHs can
  be non-spinning if the binary separations are not sufficiently small.

We calculate the spin evolution of a WR star just after
  common envelope evolution, taking into account tidal spin-up by its
  companion star, and stellar wind mass loss of the WR star itself. We
  use the simple model adopted by \cite{2017ApJ...842..111H} and
  \cite{2020ApJ...892...64P}. We outline the model here. We define the
  stellar dimensionless spin, projected onto the binary angular momentum
  vector, as
  \begin{equation}
    \chi = \frac{c}{GM^2} \boldsymbol{S} \cdot \frac{\boldsymbol{L}}{|\boldsymbol{L}|},
  \end{equation}
  where $c$, $G$, $M$, $\boldsymbol{S}$ and $\boldsymbol{L}$ are the speed of
  light, the gravitational constant, a stellar mass, a stellar spin
  angular momentum vector, and a binary angular momentum vector,
  respectively. The dimensionless spin of a WR star, $\chi_*(t)$,
  evolves as follows:
  \begin{equation}
    \frac{d\chi_*(t)}{dt} = \frac{(\chi_{\rm
        syn}-\chi_*(t))^{8/3}}{t_{\rm syn}}-\frac{\chi_*(t)}{t_{\rm
        w}}.
    \label{eq:chi_evo}
  \end{equation}
  The first and second terms on the right-hand side represent the
  effects of the tidal spin-up and the stellar wind,
  respectively. Here, $t_{\rm syn}$ is the timescale for the WR star
  to become tidally synchronized (in the absence of winds), and it is given by
  \begin{equation}
    t_{\rm syn} \sim 10 q^{-1/8} \left( \frac{1+q}{2q} \right)^{31/24}
    \left( \frac{t_{c}}{1~{\rm Gyr}} \right)^{17/8}~{\rm
      Myr}, \label{eq:SynchronizationTime}
  \end{equation}
  where $q$ is the mass ratio of the WR star to its companion. The
  timescale $t_{c}$ is the merger time of the binary through
  gravitational wave radiation:
  \begin{equation}
    t_{c} \sim 10 \left( \frac{2q^2}{1+q} \right) \left(
    \frac{a}{44R_{\odot}} \right)^4 \left( \frac{M_*}{30M_{\odot}}
    \right)^{-3}~{\rm Gyr}, \label{eq:GravitationalWaveRadiationTime}
  \end{equation}
  where $a$ is the semi-major axis of the binary, and $M_*$ is the WR
  mass. We suppose that the binary has a circular orbit, since it
  experiences common envelope evolution. Tides drive the stellar spin towards the synchronization value
  $\chi_{\rm syn}$, expressed as
  \begin{align}
    \chi_{\rm syn} &\sim 0.5 q^{1/4} \left( \frac{1+q}{2} \right)^{1/8}
    \left( \frac{\epsilon}{0.075} \right) \left(
    \frac{R_*}{2R_{\odot}} \right)^2 \left( \frac{M_*}{30M_{\odot}}
    \right)^{-13/8} \left( \frac{t_c}{1~{\rm Gyr}} \right)^{-3/8}, \\
    \epsilon &= \frac{I_*}{M_*R_*^2},
  \end{align}
  where $R_*$ and $I_*$ are the radius and inertia of the WR star,
  respectively. In this work, we set $\epsilon$ to be 0.15. The timescale of stellar wind, $t_{\rm w}$, is another parameter which we have to take into account.  The WR
  star loses its spin angular momentum through stellar winds with this timescale. We
  assume that $t_{\rm w}$ corresponds to
  the spin loss rate;
\begin{equation}
    t_{\rm w} = \frac{\chi_{*}(t)}{\dot{\chi_{*}}(t)}.
\end{equation}

The WR star finally collapses to the primary or secondary BHs
  in a BBH, respectively. We can obtain the effective spin of the BBH,
  such that
  \begin{equation}
    \chi_{\rm eff} = \frac{M_1 \chi_{1} + M_2 \chi_{2}}{M_1+M_2},
  \end{equation}
  where $M_1$ and $M_2$ are the masses of the primary and secondary
  BHs, respectively, and $\chi_1$ and $\chi_2$ are the dimensionless
  spins of the primary and secondary BHs.

We use two different models for the distribution of initial
stellar dimensionless spins just after common envelope
evolution. First, we assume that initial
  dimensionless spins are zero, i.e. $\chi_*(0)=0$ for all
stars.  The dimensionless spins of BHs
($\chi_1$ and $\chi_2$) are obtained only
by integrating equation (\ref{eq:chi_evo}) as they evolve from
WR stars to BHs.  Results in this case are shown in
Sections \ref{results1} and \ref{results2}. In Section
\ref{sec:flat_isotropic}, we also show the distribution of $\chi_{\rm
  eff}$ in the case that the distribution of initial
  dimensionless spins
follows a flat isotropic model, which means that the magnitude of
the stellar dimensionless spin just after the common envelope is assumed to be
randomly flat between 0 and 1 and the direction of the spin angular
momentum vector is isotropic. We set $t_{\rm w}$ to be 1.0 Myr as
our standard model. We also perform simulations with $t_{\rm w}=0.5$ and 3
Myr in Section \ref{sec:flat_isotropic} and compare the distribution
of the effective spins for each model.

We do not take into account tidal interactions during MS and
  post MS phases, since they lose a large fraction of their spin
  angular momenta through common envelope evolution. We do not
  consider that BHs get spin angular momenta, accreting masses from
  their companions, because BHs accrete little mass in our simulations. We assume that WR stars do not lose spin
  angular momenta.

%%%%%%%%%%%%%%%%%%%%%%%%%%%%%%%%%%%%%%%%%%%%%%%%%%
%%%%%%%%%%%%%%%%%%%%%%%%%%%%%%%%%%%%%%%%%%%%%%%%%%

\section{Results and Discussion}

\subsection{Relation between spin and BBH properties}
\label{results1}

We obtained in total $\sim$300--1000 BBHs ejected from clusters for
four different metallicity models, however, the number of BBHs merged
within 14\,Gyr (hereafter, merging BBHs) is only 7--37 for each
metallicity model. Here, we assume that $\chi_*(0)=0$ for all stars,
and calculate the evolution of effective spins for these
merging BBHs.

Table \ref{tab:BBHs} shows the number of all merging BBHs and
those with $\chi_1 > 0.1$, $\chi_2 > 0.1$, and
  $\chi_{\rm eff}>0.1$ for each metallicity. For Model Z002, there is
no BBH with $\chi_1 > 0.1$, $\chi_2 > 0.1$ or $\chi_{\rm
    eff}>0.1$.  This is because merging BBHs are formed
  through three-body interactions after BH formation (See
Paper II). As described in Section \ref{sec:Spins}, we
  consider that only BH progenitors can be tidally spun up by their
  companions just after common envelope phases. After they evolve to BHs, they can not be
  tidally spun up by their companions.

\begin{table*}
 \centering
 \caption{Number of BBHs.}
 \label{tab:BBHs}
 \begin{tabular}{lcccc}
  \hline
              & $N_{\rm mBBH}^{a}$ & $N_{\rm mBBH}(\chi_1>0.1)$ & $N_{\rm mBBH}(\chi_2>0.1)$ & $N_{\rm mBBH}(\chi_{\rm eff}>0.1)$ \\
  \hline
  Model Z0002 & 37 & 1 & 2 & 2 \\
  Model Z0005 & 17 & 2 & 3 & 3 \\
  Model Z001  & 32 & 1 & 7 & 7 \\
  Model Z002  &  7 & 0 & 0 & 0 \\
  \hline
  \multicolumn{5}{l}{\footnotesize$^a$ number of merging BBHs}\\
 \end{tabular}
\end{table*}

\begin{figure}
	\includegraphics[width=\columnwidth]{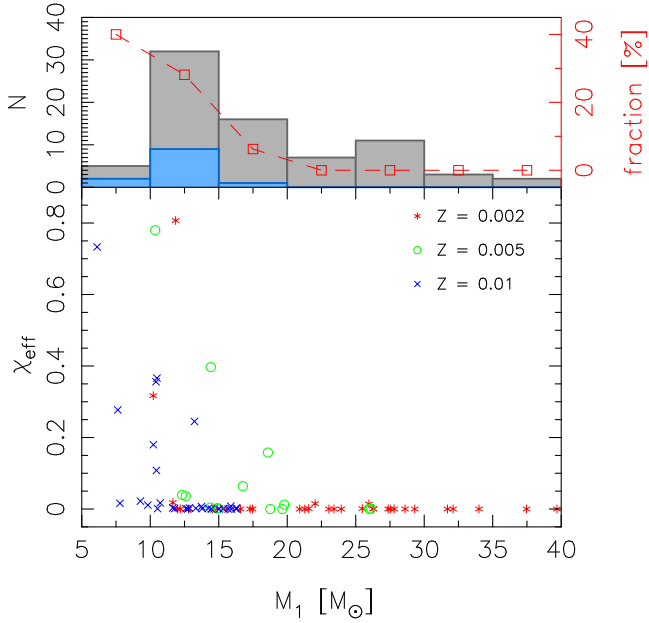}
    \caption{The bottom panel shows the distribution of the effective
      spins as a function of the primary mass of BBHs
      in the case of the zero initial dimensionless spins model. The
      different point types indicate different
      metallicities. Histograms in the top panel show the number of
      merging BBHs in each primary mass bin. The width of the bins is
      equal to $5M_\odot$. Grey and blue histograms show the number of
      BBHs with smaller and larger effective spins than
      0.1. The red line with square points shows the fraction
      of BBHs with $\chi_{\rm eff} > 0.1$ in each bin.}
    \label{fig:M1Xeff}
\end{figure}

\begin{figure}
	\includegraphics[width=\columnwidth]{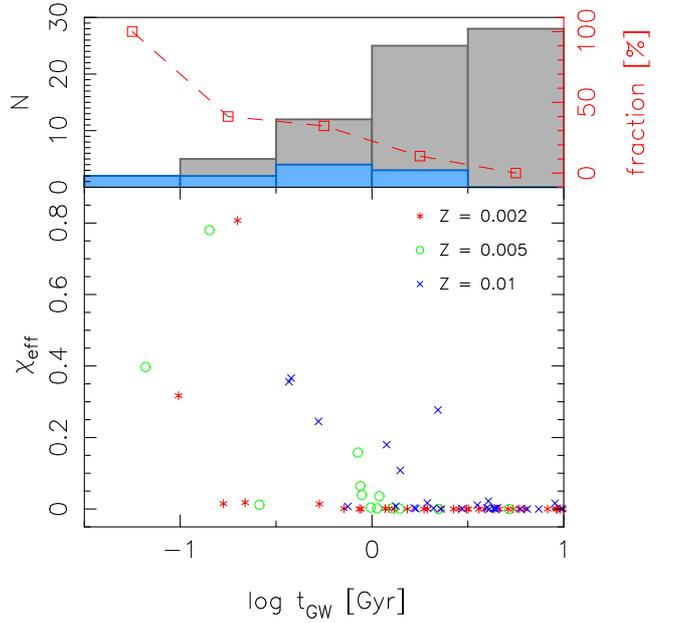}
    \caption{Same as Figure \ref{fig:M1Xeff}, but as a function of
      BBH merger time through gravitational wave radiation.}
    \label{fig:tgwXeff}
\end{figure}

\begin{figure}
	\includegraphics[width=\columnwidth]{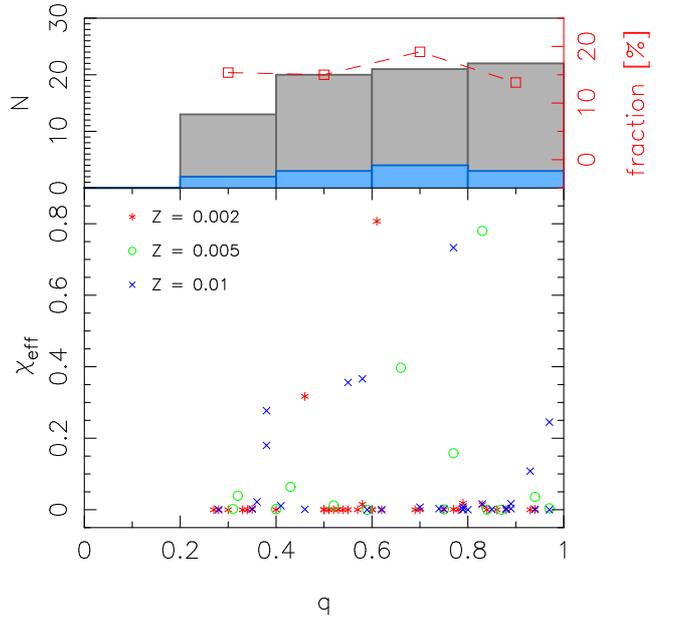}
    \caption{Same as Figure \ref{fig:M1Xeff}, but as a function of
      mass ratios of BBHs.}
    \label{fig:qXeff}
\end{figure}

Bottom panels in Figure \ref{fig:M1Xeff}, \ref{fig:tgwXeff} and
\ref{fig:qXeff} show the distribution of $\chi_{\rm eff}$ as a
function of $M_1$, $t_{\rm GW}$ (BBH merger time through
  gravitational wave radiation) and $q$ for each metallicity. Grey
and blue histograms in the top panel show the number of BBHs with
smaller and larger effective spins than 0.1. Red
  dashed lines with red square points in top panels show the fraction
of BBHs with $\chi_{\rm eff} > 0.1$ in each histogram bin. 

Note that the number of simulation runs is different for each metallicity model. Consequently, the spin distribution does not reflect the actual distribution of merging BBHs in the local Universe. This result only describes a rough trend in the spin distribution.

In Figure \ref{fig:M1Xeff}, our results show that only less massive
BBHs ($M_1\lesssim 20M_\odot$) have larger effective spins than
0.1. This result is interpreted as follows. Lower-mass BHs
  form from lower-mass WR stars. Such lower-mass WR stars
  are in binaries with smaller semi-major axes after common envelope
  evolution, since binaries with lower-mass members need smaller
  binding energies (in absolute value) to shrink. WR stars in binaries
  with smaller semi-major axes are more tidally spun up as seen in
  equations (\ref{eq:SynchronizationTime}) and
  (\ref{eq:GravitationalWaveRadiationTime}).

A BBH with a smaller semi-major axis has a shorter merger
time. This is consistent with the results shown in Figure
\ref{fig:tgwXeff}, which shows that BBHs with shorter merger times
tend to have larger spins. Figure \ref{fig:qXeff} show that the
fraction of BBHs with $\chi_{\rm eff} > 0.1$ is independent on the
mass ratio.

\subsection{Spin distribution with the zero initial dimensionless spin}
\label{results2}

\begin{figure}
	\includegraphics[width=\columnwidth]{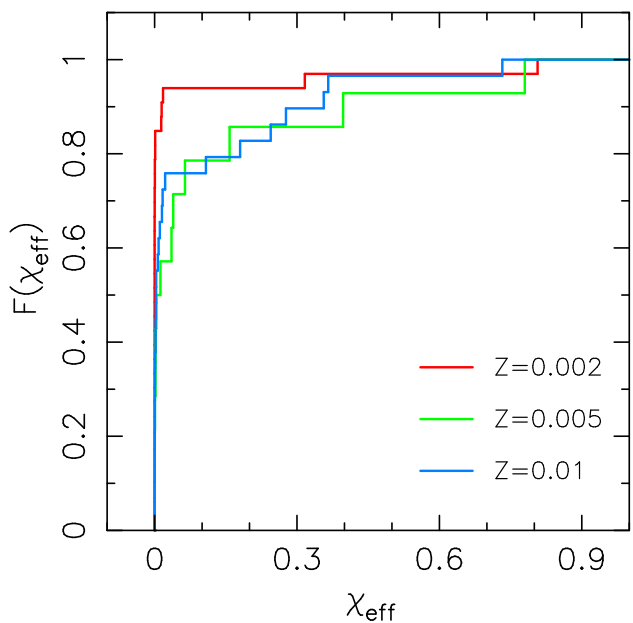}
    \caption{Cumulative distribution of effective spins of
      merging BBHs for each metallicity model in the case of the zero
      initial dimensionless spins model.}
    \label{fig:Xcdf_Z}
\end{figure}

Figure \ref{fig:Xcdf_Z} shows the cumulative distribution of effective
spins, $F(\chi_{\rm eff})$, for each metallicity model. For
the 0.1 solar metallicity model (red line), the effective spin is
almost zero for more than 90\% of merging BBHs. For the 0.25 (green
line), 0.5 (blue line) solar metallicity models, over 20\% of
the merging BBHs show larger effective spins than 0.1.

BBHs formed in more metal-rich clusters tend to get
larger effective spins, since they have lower-mass BHs. For
the solar-metallicity clusters, however, the effective
spins of all merging BBHs are almost zero. This is
  because most of these BBHs experienced a three-body interaction and
  shrank their orbits to become merging BBHs (See Paper II). Although
  some of BBHs are formed via common envelope evolution, the resulting
  BBH progenitors have too large separations to be spun up through
  tidal interactions. For the solar-metallicity clusters, stars lose a
  large part of their envelopes due to strong stellar winds before
  they experience common envelope evolution.

In order to calculate the local merger rate density, we have obtained
the cosmic star formation history for our four metallicity model from
the observation \citep{2019MNRAS.488.5300C} in Paper II (See also
Figure 8 of Paper II). Using the cosmic star formation history for
each metallicity, we estimate the local spin distribution of merging
BBHs.

\begin{figure}
	\includegraphics[width=\columnwidth]{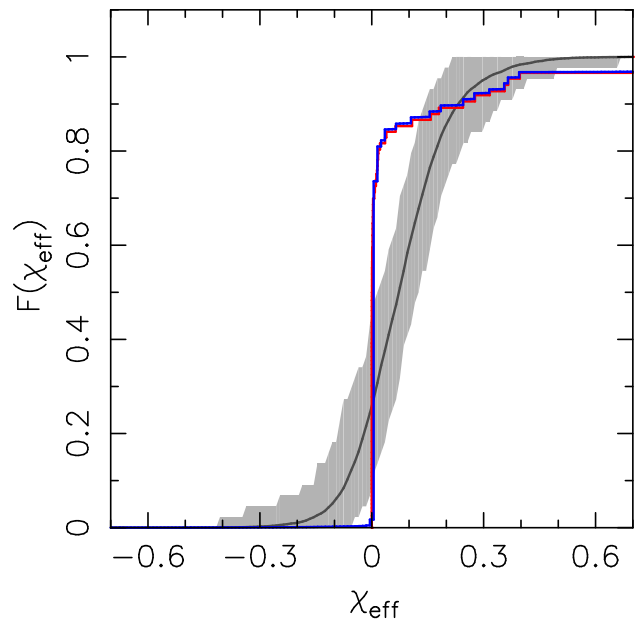}
    \caption{The cumulative local distribution of the effective
      spins in the case of the zero initial dimensionless spins model.
      The red line shows the result obtained from our
      simulations. The blue line indicates the effective spin
      distribution, taking into account spin tilt by single
      encounters after BBH formations. The solid black line shows
      the one estimated from gravitational wave detection 
      \citep[see also Figure 24. of][]{2020arXiv201014533T}. The grey 
      region shows the central 98\% credible bounds on the posterior
      predictive distributions.}
    \label{fig:Xcdf}
\end{figure}

The red line in Figure \ref{fig:Xcdf} shows the local spin
distribution $F(\chi_{\rm eff})$ obtained from our simulation. The
grey region shows the central 98\% credible bounds on the posterior
predictive distributions \citep{2020arXiv201014533T}. Our results show
that most of merging BBHs in the local Universe have zero
effective spins. In these calculations, we assume that the
initial dimensionless spins of WR stars just after
common envelope evolution are zero.
In Section \ref{sec:flat_isotropic}, we consider WR stars
  which have non-zero spins tilt from the binary angular momentum
  vectors just after common envelope evolution.

We obtain the merger rate of BBHs with large
  effective spins. From the cumulative distribution in
Figure \ref{fig:Xcdf},
Therefore, about 16\% of merging BBHs have effective spins
  larger than 0.1 in the local Universe. In Paper II, we have estimated that the
local merger rate density of BBHs originated from the open clusters is
about $70~{\rm yr}^{-1}{\rm Gpc}^{-3}$. Thus, the local merger
rate density of BBHs with larger $\chi_{\rm eff}$ than 0.1 is
\begin{equation}
    R_{\chi_{\rm eff}>0.1} \sim 11~{\rm yr}^{-1}{\rm Gpc}^{-3}.
\end{equation}
It might seem odd that the local merger rate density of
  all the BBHs and BBHs with $\chi_{\rm eff}>0.1$ are, respectively,
  more than and comparable to the local merger rate density of all the
  BBHs inferred by gravitational observations
  \citep{2020arXiv201014533T}. Rather, our predicted merge rate
  density should be consistent with the observed one. This is because
  we ignore BH natal kicks via asymmetric supernova explosions, which
  decreases the local merger rate density of BBHs by several factors
  \citep{2017PASJ...69...94F}.

\begin{figure}
	\includegraphics[width=\columnwidth]{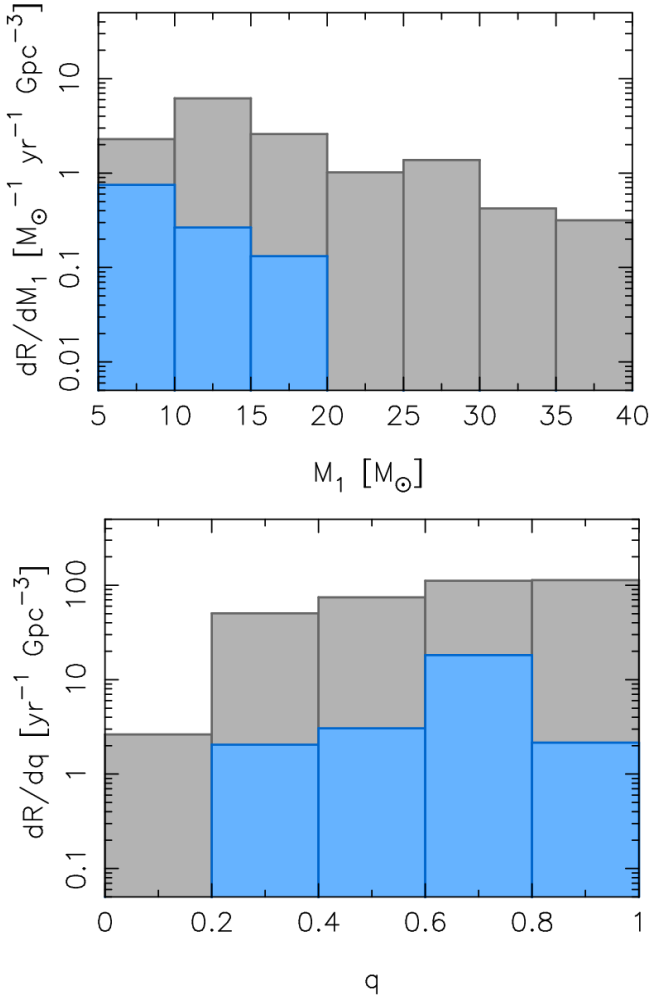}
    \caption{Differential merger rate density for primary BH
      masses, $M_1$, (top panel) and mass ratios, $q$,
      (bottom panel) in the case of the zero initial dimensionless
      spins model. Grey and blue histograms show the differential
      merger rate density of BBHs with effective spins smaller
        and larger than 0.1, respectively.}
    \label{fig:dRdM1}
\end{figure}

Figure \ref{fig:dRdM1} shows the differential merger rate density for
primary BH masses, $M_1$, (top panel) and mass
  ratios $q$ (bottom panel). The histograms with blue and
grey areas are the differential local merger rate density
expected from our simulations (the same as Figure 9 in Paper
II). The blue area is the differential local merger rate density when
we focus only on BBHs with $\chi_{\rm eff}>0.1$. As shown in Figure 2,
BBHs with lower primary masses tend to have larger
spins.

Finally, we investigate the effect of dynamical interactions
  of merging BBHs with other BHs. As \cite{2021arXiv210201689T} found,
  ${\sim}10\%$ of merging BBHs with moderately spinning BHs can
  experience single dynamical interactions with other BHs in open
  clusters until they merge, or are ejected from open clusters. Then,
  such interactions should cause spin-orbit misalignments, and change
  the effective spin distribution. Using the misalignment distribution
  of the case with open cluster (OC) and $Z=0.002$ (see the green dashed curve in the
  middle panel of Fig. 2 in \cite{2021arXiv210201689T}), we modify the
  effective distribution as seen in the blue curve in Figure
  \ref{fig:Xcdf}. Since there are many BBHs with $\chi_{\rm eff} = 0$,
  the distribution is still not in good agreement with the observed
  one. However, the distribution may be more probable. This is because
  BBHs with negative $\chi_{\rm eff}$ appear, and there are still more
  BBHs with positive $\chi_{\rm eff}$ than with negative $\chi_{\rm
    eff}$.

\subsection{Spin distribution with the flat isotropic model}
\label{sec:flat_isotropic}

Figure \ref{fig:Xcdf} shows that $\chi_{\rm eff}$ is equal to zero for
many BBHs. This is because initial dimensionless spins was
assumed to be zero for all WR stars just after common
envelope evolution. Therefore, all BBHs without
tidally spun up show $\chi_{\rm eff}=0$. In this subsection, we
assume the flat isotropic model for the initial dimensionless
spins (see section \ref{sec:Spins}).
We denote $\chi_{*}(0)$ of primary and secondary WR stars by
$\chi_{*,1}(0)$ and $\chi_{*,2}(0)$, respectively. 20 samples of
$\chi_{*,1}(0)$ and $\chi_{*,2}(0)$ are randomly assigned to each binary
formed in the simulation, and the final $\chi_{\rm eff}$ is
calculated. For the BBHs dynamically assembled from isolated BHs,
which do not experience common envelope evolution, we assume
that the stellar dimensionless spins are uniformly distributed between 0 and 1 at
the formation time, and calculate the subsequent spin
evolution. Naturally, the tidal spin-up effect does not affect isolated WR stars, and only the spin loss due to the stellar wind is
calculated. We also assume that the spin angular momentum vector is
randomly oriented when the binary is formed by dynamical interaction.

\begin{figure*}
	\includegraphics[width=0.96\textwidth]{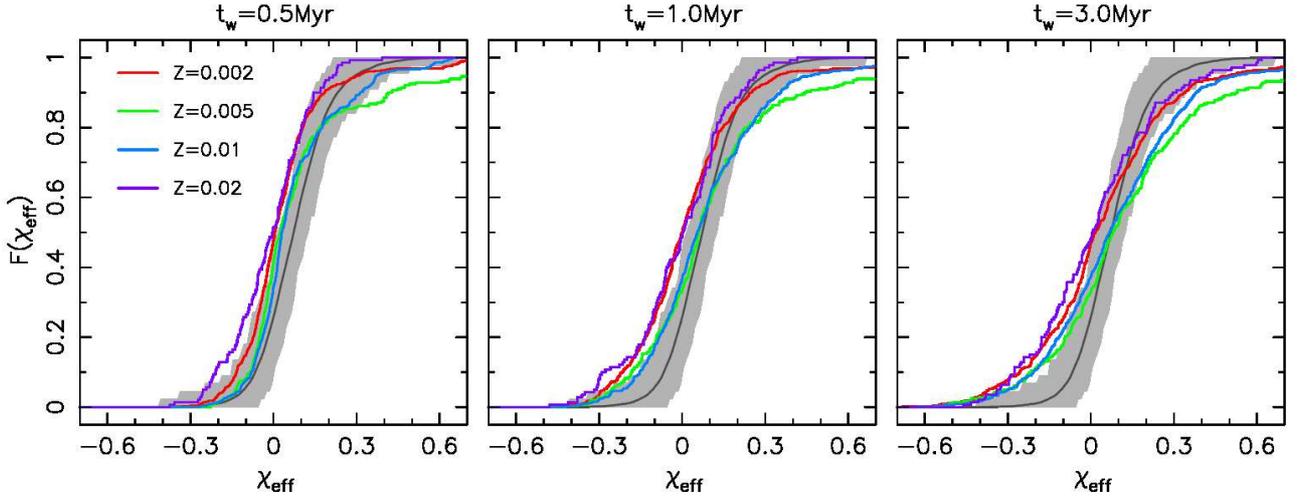}
    \caption{Same as Figure \ref{fig:Xcdf_Z}, but in the case of the
      flat isotropic model for $\chi_*(0)$ with $t_{\rm w}=$0.5, 1.0
      and 3.0 Myr from left to right panel.}
    \label{fig:Xcdf_Z2}
\end{figure*}

Figure \ref{fig:Xcdf_Z2} shows the cumulative distribution of
$\chi_{\rm eff}$ of merging BBHs with the flat isotropic model for
the initial dimensionless spins with $t_{\rm w}=$ 0.5, 1.0 and
3.0 Myr. In the case of shorter $t_{\rm w}$, more spin angular
momentum is lost during the evolution from WR to BH than the case of
longer $t_{\rm w}$. Therefore, for models with smaller $t_w$,
$\chi_{\rm eff}$ is more concentrated around zero.

\begin{figure}
	\includegraphics[width=\columnwidth]{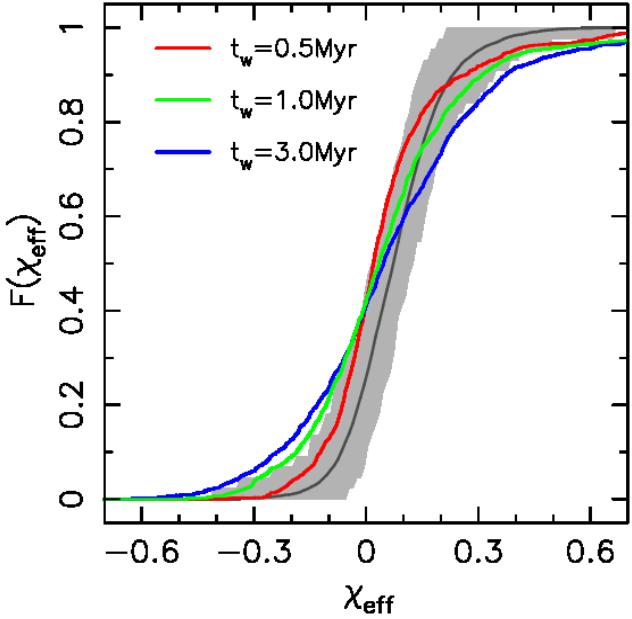}
    \caption{Same as Figure \ref{fig:Xcdf}, but in the case of the
      flat isotropic model for $\chi_*(0)$ with $t_{\rm w}=$0.5, 1.0
      and 3.0 Myr.}
    \label{fig:Xcdf2}
\end{figure}

In Figure \ref{fig:Xcdf2}, we present the same plot as Figure \ref{fig:Xcdf},
but in the case of the flat isotropic model for the initial
  dimensionless spins with $t_{\rm w}=$ 0.5, 1.0 and 3.0 Myr. By
considering the flat isotropic model for the initial
  dimensionless spins, the $\chi_{\rm eff}$ is distributed around
zero and slightly biased to the positive side due to the effect of the
tidal spin-up. This distribution is similar to the one
estimated from gravitational wave observation
\citep{2020arXiv201014533T}.

\begin{figure}
	\includegraphics[width=\columnwidth]{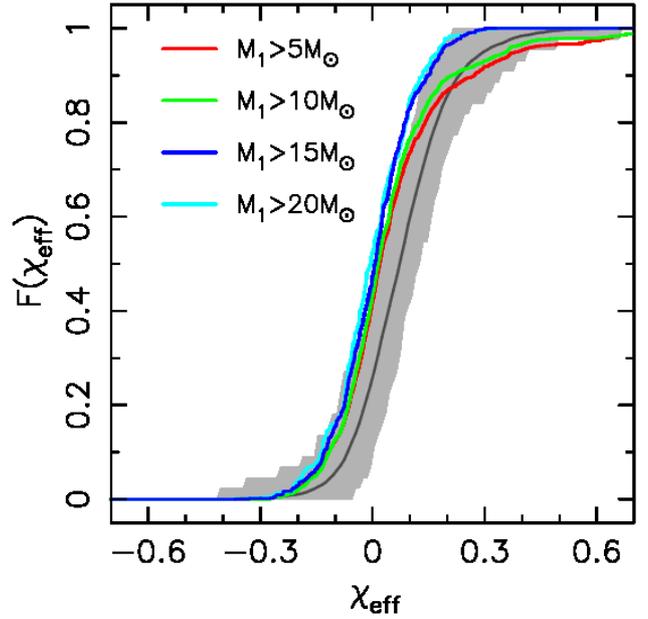}
    \caption{Same as case of $t_{\rm w}=$0.5 Myr in Figure
      \ref{fig:Xcdf2}, but results obtained from different primary
      masses range.}
    \label{fig:Xcdf3}
\end{figure}

We investigate the dependence of the cumulative distribution
on the primary mass.  Figure \ref{fig:Xcdf3} shows the cumulative
distribution of $\chi_{\rm eff}$ for the merging BBHs with some
primary mass ranges. BBHs with smaller primary masses
have larger effective spins.
This is because BBHs with smaller primary masses tend to have
larger effective spins, as shown in Figure 2. This
may be consistent with a negative correlation between BBH masses and
effective spins which \cite{2020ApJ...894..129S} have shown, using
the data of the first and second observing runs of LIGO and Virgo.

%%%%%%%%%%%%%%%%%%%%%%%%%%%%%%%%%%%%%%%%%%%%%%%%%%
%%%%%%%%%%%%%%%%%%%%%%%%%%%%%%%%%%%%%%%%%%%%%%%%%%

\section{Conclusions}

We performed direct N-body simulations of open clusters with a mass of
$2500~M_{\odot}$ and four different metallicity models ($Z=0.1, 0.25,
0.5, 1.0~Z_{\odot}$). We investigated the spin evolution of WR
stars tidally spun-up by their close companions,
which evolve to BBHs as the sources of gravitational
waves. As a result, we found the following:
\begin{enumerate}
\item Some of the BBHs with small semi-major axes
  have greater effective spins than 0.1 because of
  strong tidal forces from the companions, even
  assuming that BH progenitors have zero spins immediately
 after common envelope evolution.
\item BBHs have larger effective spins in more metal-rich
  clusters except for the solar-metallicity clusters. More metal-rich
  clusters form lower-mass BHs due to stellar wind mass loss, and
  lower-mass BHs have tighter WR stars just after common envelope
  evolution. The WR stars are more tidally spun-up by the BHs. In the
  solar-metallicity clusters, most of merging BBHs are formed not
  through common envelope evolution, but through dynamical
  interactions.
\item We estimated the effective spin distribution of merging BBHs
  formed in open clusters as gravitational wave sources. We
  found that about 16\% of local merging BBHs have 
    effective spins larger than 0.1. Hence, the local merger rate density of
  BBHs with $\chi_{\rm eff}$ larger than 0.1 is $\sim 11~{\rm
    yr}^{-1}{\rm Gpc}^{-3}$, if we adopt $\sim 70~{\rm
      yr}^{-1}{\rm Gpc}^{-3}$ for the local merger rate density of all
    the BBHs as described in Paper II.
\item Considering the flat isotropic distribution of
    dimensionless spins for WR stars just after common
    envelope evolution, the distribution of $\chi_{\rm eff}$
    extends around zero and is slightly biased to the positive side
    due to the effect of tidal spin-up of BH progenitors.
    This distribution is similar to the one inferred from gravitational wave
    observations \citep{2020arXiv201014533T}.
\end{enumerate}

Future observations of gravitational waves, which will provide information
on the spin of BBHs, especially those of small masses, will allow us
to discuss in detail the spin evolution mechanism of BBHs in open
clusters.

%%%%%%%%%%%%%%%%%%%%%%%%%%%%%%%%%%%%%%%%%%%%%%%%%%
%%%%%%%%%%%%%%%%%%%%%%%%%%%%%%%%%%%%%%%%%%%%%%%%%%

\section*{Acknowledgments}

This work is supported by JSPS KAKENHI Grant Number 17H06360,
19H01933, and 19K03907 and The University of Tokyo Excellent Young
Researcher Program, and MEXT as “Program for Promoting Researches on
the Supercomputer Fugaku” (Toward a unified view of the universe: from
large scale structures to planets, Revealing the formation history of
the universe with large-scale simulations and astronomical big
data). Numerical calculations reported in this paper were supported by
Initiative on Promotion of Supercomputing for Young or Women
Researchers, Supercomputing Division, Information Technology Center,
The University of Tokyo.


\bibliographystyle{mnras}
\bibliography{main} % if your bibtex file is called example.bib

%%%%%%%%%%%%%%%%%%%%%%%%%%%%%%%%%%%%%%%%%%%%%%%%%%

% Don't change these lines
\bsp	% typesetting comment
\label{lastpage}
\end{document}